\documentclass[12pt,preprint]{aastex}

\newif\ifAMStwofonts                        
\newcommand{\lsimeq}{{_<\atop^{\sim}}}

\shorttitle{SEDs of MIR sources}
\shortauthors{Gruppioni et al.}

\begin{document}

\title{The contribution of AGN and star-forming galaxies to the Mid-Infrared as revealed by their Spectral Energy Distributions}
\author{C. Gruppioni\altaffilmark{(1)}$^a$, F. Pozzi\altaffilmark{(2)}, M. Polletta\altaffilmark{(3,4)}, G. Zamorani\altaffilmark{(1)}, F. La Franca\altaffilmark{(5)}, N. Sacchi\altaffilmark{(5)}, A. Comastri\altaffilmark{(1)}, L. Pozzetti\altaffilmark{(1)}, C. Vignali\altaffilmark{(2)}, C. Lonsdale\altaffilmark{(6)}, M. Rowan-Robinson\altaffilmark{(7)}, J. Surace\altaffilmark{(6)}, D. Shupe\altaffilmark{(6)}, F. Fang\altaffilmark{(6)}, I. Matute\altaffilmark{(8)}, S. Berta\altaffilmark{(9)}}

\altaffiltext{(1)}{INAF: Osservatorio Astronomico di Bologna, via Ranzani 1, I--40127 Bologna, Italy\\
$^a$ e-mail: carlotta.gruppioni@oabo.inaf.it}
\altaffiltext{(2)}{Dipartimento di Astronomia, Universit\`a di Bologna, via Ranzani 1, I--40127 Bologna, Italy}
\altaffiltext{(3)}{Institut d'Astrophysique de Paris, 98bis blvd Arago, Paris 75014, France}
\altaffiltext{(4)}{INAF: IASF Milano, via Bassini 15, I--20133 Milano, Italy}
\altaffiltext{(5)}{Dipartimento di Fisica, Universit\`a degli Studi ``Roma Tre'', via della Vasca Navale 84, I--00146 Roma}
\altaffiltext{(6)}{Infrared Processing and Analysis Center, California Institute of Technology 100-22, Pasadena, CA 91125, USA}
\altaffiltext{(7)}{Astrophysics Group, Blackett Laboratory, Imperial College of Science Technology and Medicine, Prince Consort Road, London SW7 2BZ, UK}
\altaffiltext{(8)}{INAF: Osservatorio Astrofisico di Arcetri, Largo E. Fermi 5, I--50125 Firenze, Italy}
\altaffiltext{(9)}{Dipartimento di Astronomia, Universit\'a di Padova, vicolo Osservatorio 2, I--35122 Padova, Italy}

\begin{abstract}
\noindent We present the broad-band Spectral Energy Distributions (SEDs) of the largest available
highly (72\%) complete spectroscopic sample of mid-infrared (MIR) selected galaxies and AGN at intermediate redshift. 
The sample contains 203 extragalactic sources from the 15-$\mu$m survey in 
the ELAIS-SWIRE field S1, all with measured spectroscopic redshift. 
Most of these sources have full multi-wavelength coverage from 
the far-UV ({\em GALEX}) to the far-infrared ({\em Spitzer}) and lie in the redshift range $0.1<z<1.3$.
Due to its size, this sample allows us for the first time to characterise 
the spectral properties of the sources responsible for the strong evolution observed in the MIR. 
Based on SED-fitting technique we have classified the MIR sources, identifying 
AGN signatures in about 50\% of them. This fraction is significantly higher than that derived 
from optical spectroscopy ($\sim$29\%) and is due in particular to the identification of 
AGN activity in objects spectroscopically classified as galaxies.   
This might be partially due to the fact that the spectroscopic classification can be somewhat unreliable
because of host galaxy dilution in the optical. 
It is likely that in most of our objects, the AGN is either obscured
or of low-luminosity, and thus it does not dominate the energetic output at any wavelength, except in the MIR, showing
up just in the range where the host galaxy SED has a minimum.
The fraction of AGN strongly depends on the flux density, with that derived through the SED-fitting being about 20\%
at $S_{15 \mu m} \sim 0.5-1$ mJy and gradually increasing up to 100\% at $S_{15 \mu m} > 10$ mJy, while that obtained from 
optical spectroscopy 
never being $>$30\%, even at the higher flux densities. The results of this work will 
be very useful for updating all the models aimed at interpreting the deep 
infrared survey data and, in particular, for constraining the nature and the role of dust-obscured 
systems in the intermediate/high-redshift Universe.
\end{abstract}


\section{Introduction}
\noindent Understanding the overall Spectral Energy Distribution (SED) of sources responsible for the observed Cosmic Infrared Background 
(CIRB) is a crucial tool for unveiling the nature and evolution of galaxies and AGN in the infrared (IR) and for obtaining a complete
picture of the history of star-formation and obscured AGN activity in the Universe. 
The mid- and far-infrared (MIR and FIR) regions of the electromagnetic spectrum
efficiently probe the population of actively star-forming galaxies and dust-obscured AGN.

\noindent The {\em InfraRed Astronomical Satellite} {\em IRAS} has sampled the local Universe ($z < 0.2$) in the MIR/FIR band, discovering Ultra Luminous Infrared Galaxies (ULIGs: $L_{IR} > 
10^{12}L_{\odot}$; Sanders \& Mirabel 1996, Lonsdale, Farrah \& Smith 2006) and first showing 
some evidence for their strong evolution (i.e. Hacking, 
Houck and Condon 1987, Lonsdale \& Hacking 1989; Saunders et al. 1990).
 
\noindent Extragalactic source counts from different surveys over a wide flux range
obtained with the {\em ISOCAM} instrument (Cesarsky et al. 1996) on board of the {\em Infrared Space
Observatory} ({\em ISO}; Kessler et al. 1996) indicate that also IR sources less extreme than ULIGs (i.e. Luminous Infrared Galaxies; LIGs: $L_{IR} > 10^{11} L_{\odot}$) have evolved rapidly from 
$z \sim 0$ up to $z \sim 1.2$, significantly faster than deduced from 
optical surveys (i.e. Elbaz et al. 1999; Gruppioni et al. 2002; Pozzi et al. 2004). 

\noindent The {\em Spitzer} Space Telescope (Werner et al. 2004) is now providing new insights into the
IR population contributing to the CIRB, in particular in the {\em MIPS} 24-$\mu$m band,
where the extragalactic source counts (Marleau et al. 2004;
Papovich et al. 2004) have confirmed the existence of the rapidly evolving dust-obscured 
population discovered by {\em ISOCAM}, detecting the higher-$z$ analogues ($1 < z < 3$) of 
the {\em ISOCAM} galaxies. In particular, the deeper {\em MIPS} surveys are now able to
resolve about 70\% of the CIRB at 24 $\mu$m, with the bulk of this background 
originating in discrete sources at $z \lsimeq 2$ (e.g. Papovich et al. 2004).

These results have stimulated the development
of several evolutionary models for IR galaxies (i.e. Rowan-Robinson 2001;
Franceschini et al. 2001; Chary \& Elbaz 2001; Xu et al. 2003; Lagache, Dole \& Puget 2003; 
Lagache et al. 2004; Pozzi et al., 2004; Gruppioni et al. 2005; Pearson 2005), aimed at reproducing the 
MIR source counts and the observed redshift distributions.
All these models are more or less able to reproduce the observed {\em ISOCAM} 15 $\mu$m source counts, 
but none of them can provide an acceptable fit to the {\em MIPS} 24-$\mu$m counts without any ``ad-hoc'' changes. 
In particular, they tend to
predict the characteristic turn-over of the differential 24-$\mu$m number counts at fluxes higher than
observed. Even worse is the situation regarding the observed redshift distributions, both for {\em ISOCAM} 
and {\em MIPS} sources.
Le Floc'h et al. (2005) and P\`erez-Gonz\`alez et al. (2005), by studying the luminosity
evolution of a sample of {\em MIPS} 24-$\mu$m sources in the Chandra Deep Field South (CDFS) up to
$z=1$ and in the range $1< z < 3$ respectively, find that all the models considered in these works 
(Lagache et al. 2004, Chary et al. 2004, Gruppioni et al. 2005 and Pearson et al. 2005) fail in reproducing 
the differential
counts at 24 $\mu$m in different redshift slices and the total $z$-distribution. We must note that
most of the redshifts considered in these works are photometric and the analysed fields are relatively small, 
and can therefore be affected by cosmic variance. However, there seems to be a significant disagreement between
all the evolutionary models existing in literature and the recent {\em Spitzer} data. 
Two main weaknesses are common to all these models:
\begin{itemize}
\item[1)] They are based on very uncertain assumptions about the shape of the  
SED of MIR sources. In particular, they generally assume an invariant PAH emission and extrapolate few 
local SEDs to high redshifts and luminosities (L), either by keeping them fixed or artificially increasing 
their activity with luminosity (e.g. Chary \& Elbaz 2001). 
\item[2)] They severely underestimate, or even completely neglect (e.g. Lagache et al. 2003; Chary \& Elbaz 2001), the 
AGN contribution in MIR sources (see Brand et al. 2006; Gruppioni et al. 2006). 
In fact, without MIR spectroscopic data or, if not available, without a complete SED characterisation, it is very difficult to disentangle star-forming galaxies from AGN (Genzel et al. 1998, Sajina et al. 2007).
This is particularly true in cases where spectroscopic classification based on optical line diagnostics is 
impossible due to the absence of some of the required emission lines in the observed spectral range, or to the significant 
differential line extinction produced by dust, or to the lack of spectroscopic data and deep X-ray coverage.
Since different evolutionary models are applied to the two populations, the misidentification of AGN and
star-forming galaxies introduces large uncertainties in the predicted source counts and backgrounds..
\end{itemize}

The crucial wavelength range (3-200 $\mu$m) covered 
by Spitzer with unprecedented sensitivities gives the opportunity to characterise for the first time the MIR/FIR 
spectral properties of large numbers of sources over a significant extent in redshift and to study their evolution 
with z and/or L, testing the model assumptions. In particular, thanks also to the extensive multi-wavelength 
coverage available in several areas of the sky, it is now possible to study the broad-band (from UV to FIR) spectral 
properties of the same sources responsible for the observed evolution in the MIR, thus constructing observational 
templates libraries over a large range of wavelengths. 

In the Southern field of the European Large Area {\em ISO} Survey (ELAIS; Oliver et 
al. 2000; Rowan-Robinson et al. 2004), S1, an extensive multi-wavelength 
follow-up campaign has been performed during the past years. This field, initially selected for {\em ISO}
observations at 7, 15 and 90 $\mu$m, is now one of the six areas covered by the {\em Spitzer} Wide-Area InfraRed 
Extragalactic Survey (SWIRE, Lonsdale et al. 2003) in all the {\em IRAC} and {\em MIPS} bands.
In addition to {\em ISO} and {\em Spitzer} observations, the whole S1 field ($\sim$4 deg$^2$) has been deeply 
surveyed in the radio ({\em ATCA}; Gruppioni et al. 1999; Ciliegi et al., in preparation), in the optical ($R$ band, {\em ESO/3.6-m}; La Franca et al. 2004) and in the
near- and far-UV [Galaxy Evolution Explorer ({\em GALEX}); Burgarella et al. 2005]. The central square degree of S1 is covered by 
NIR (ESO/SOFI; Matute et al. in preparation) and optical ({\em ESO/WFI}; Berta et al. 2006) data, while $\sim$0.6 deg$^2$ have been observed in the
X-rays ({\em Beppo-SAX}; Alexander et al. 2001 and {\em XMM-Newton}; Puccetti et al. 2006). Optical spectroscopic 
informations are available for $\sim$200 15-$\mu$m ISOCAM sources ({\em ESO/3.6-m}; La Franca et al. 2004) and for most of the {\em XMM} and 
some of the SWIRE-24$\mu$m sources having optical counterpart brighter than 
$R \approx 24$ (with {\em ESO/VIMOS}; Feruglio et al., 2008; La Franca et al., in preparation). 

Due to its extensive multi-wavelength coverage and its large area extent, S1 is well suited to study 
in detail the broad-band SEDs of a statistically significant sample of infrared galaxies and AGN at intermediate redshifts ($z < 1.5$).
To this purpose, we have selected the 203 extragalactic sources detected by {\em ISOCAM} at 15 $\mu$m ($S > 0.5$ mJy) in the S1 field 
(Lari et al. 2001) with $R < 23$, available spectroscopic redshift and classification (La Franca et al. 2004; La Franca et al. 2007). 
The same sample has previously been used to derive the first 15-$\mu$m
luminosity function of galaxies (Pozzi et al. 2004) and AGN (Matute et al. 2002, 2006). The treatment of  
infrared/optical incompleteness of this sample for statistical purposes is described in detail by Gruppioni et al. (2002) and La Franca et al. (2004). 
The sample discussed here can be considered as the largest available spectroscopic sample of IR galaxies and AGN at intermediate redshift, with such a high level of completeness 
(72\%), allowing us for the first time to characterise the spectral properties of the 
sources responsible 
for the strong evolution observed in the MIR. Although large spectroscopic samples are available 
in fields observed by {\em Spitzer}, like the GOODS-CDFS (see Vanzella et al. 2005; 2006),
the targets are not selected to be MIR sources, but optical/NIR objects responding to given
colour criteria and likely to be at high redshifts. In the very near future large 
spectroscopic samples of MIR (24-$\mu$m) selected sources, either locally (i.e. in the FLS by Marleau et al. 2007) or at high redshift (i.e. in the COSMOS area) will be made available to the community,
allowing studies similar to the one presented here, but at present our sample is unique to this purpose.
Indeed most of our objects are LIGs, 
with MIR fluxes in the crucial range between the {\em IRAS} surveys (200 mJy) and the Deep {\em ISOCAM} Surveys (0.1 mJy; Elbaz et al. 1999), 
where MIR source counts start diverging from no-evolutionary expectations. Most of these objects (except for type 1 AGN) are
in the redshift range $0.1<z<1.0$. We make use of all the available data, 
from FUV to FIR, to derive the SEDs of these sources and construct the first
observational library of templates for MIR galaxies and AGN at intermediate $z$. To interpret the observed 
SEDs we perform a fit 
with several local 
template SEDs, representative of different classes of IR galaxies and AGN (Polletta et al. 2007), comparing the resulting
SED classification with the spectroscopic one.

The present paper is structured as follows. In Section \ref{sample} we describe the reference sample. In Section \ref{sed} we
present the multi-wavelength data-set and the observed SEDs, discussing the SED-fitting procedure and results. In Section \ref{class}
we compare the original spectroscopic classification with the SED classification. In Section \ref{discuss} we present the revised 15-$\mu$m
counts for AGN and galaxies, we discuss the results and present our conclusions.

Throughout the paper we adopt a cosmological model with $\Omega_m=0.3$, $\Omega_{\Lambda}=0.7$ and $H_0=70$ km s$^{-1}$ Mpc$^{-1}$.
The magnitudes discussed here are in the Vega system.  

\section{The Sample}
\label{sample}
Our reference sample is the Lari et al. (2001) catalogue of 462 sources selected at 15 $\mu$m (LW3 band) in the ELAIS Southern field S1
 (centred at 
$\alpha$(2000) = 00$^h$ 34$^m$ 44.4$^s$, $\delta$(2000) = -43$^{\circ}$ 28$^{\prime}$ 12$^{\prime \prime}$, and covering about $2^{\circ} 
\times 2^{\circ}$). This sample, complete at the 5$\sigma$ level, is the only {\em ISOCAM} sample covering the whole flux density range 
0.5 -- 150 mJy, thus linking {\em IRAS} to the Deep ISOCAM Surveys (Elbaz et al. 1999). The source counts at 15 $\mu$m obtained from 
that catalogue sample the flux density region where observed counts start diverging from no evolution models, as discussed by 
Gruppioni et al. (2002). 
La Franca et al. (2004) presented R-band data for a highly reliable sub-sample of 406 out of the total 462 15-$\mu$m 
sources of the Lari et al. (2001) catalogue. The R band data were obtained by an ESO imaging campaign with the {\em DFOSC} instrument 
mounted on the 1.5-m ESO/Danish telescope at La Silla (Chile), providing a reliable optical counterpart down to
$R \sim 23$ for 317 of them, thus reaching 95\% completeness level.
Spectroscopic observations of the optical counterparts of the {\em ISOCAM} S1 sources were carried out at the {\em AAT/2dF}, 
ESO/Danish 1.5-m, 
3.6-m and NTT telescopes (La Franca et al. 2004), providing a secure spectroscopic identification for 290 {\em ISOCAM} sources 
(72\% of the whole highly reliable ELAIS-S1 sample). La Franca et al. (2004) have classified 199
of the spectroscopically identified sources as extragalactic objects (25 type 1 AGN, 23 type 2 AGN, 9 liners, 32 
starburst galaxies, 100 H$\alpha$ emitter galaxies, 3 early type galaxies, 7 unclassified, but with measured redshift), the
remainder 91 as stars. 
Details about the identification completeness of the extragalactic sample for statistical uses are given in La Franca et al. (2004). 
These 199 extragalactic objects with measured redshift are the same objects previously used for the statistical analysis of the 
evolution of galaxies and AGN by Pozzi et al. (2004) and Matute et al. (2002, 2006) respectively. With respect to the La Franca et al. (2004) catalogue, two more 15 $\mu $m sources were spectroscopically identified through VIMOS-VLT observations (La Franca et al. 2007):  ELAISC15$\_$J003317$-$431706 is a $R=24.3$ galaxy showing [OII] emission at redshift 0.689, while ELAISC15$\_$J003447$-$432447 is an AGN2 at redshift 1.076. Moreover, we derived redshifts for two additional 15 $\mu$m sources: for ELAISC15$\_$J003915$-$430426 we found a $z$ value of 0.013 in the NED database, while for ELAISC15$\_$J003545$-$431833 we were able to measure $z$ through a more accurate reduction of the spectrum. Three sources with previously poor quality spectra changed their spectroscopic classification after having been re-observed with VIMOS at ESO-VLT (La Franca et al. 2007): ELAISC15$\_$J003330$-$431553, which was wrongly classified as a starburst galaxy at $z=0.473$, showed broad CIII and MgII emission at $z=2.170$ typical of AGN1 activity; ELAISC15$\_$J003603$-$433155, which was classified as AGN2, showed a broad MgII emission typical of AGN1 activity; and ELAISC15$\_$J003622$-$432826, which was classified as a starburst galaxy, showed a clear [OIII]/H$\beta$ ratio typical of AGN2 activity.  

In summary, we have considered 203 extragalactic sources selected at 15 $\mu$m with $R < 23$, spectroscopic redshift and multi-band photometry.
 
\section{The SEDs}
\label{sed}
\subsection{The Multi-Wavelength Data-Set}
\label{multil}
The objects we focus on cover the redshift range 0.01 -- 3.09 (0.01 -- 1.30 
if excluding type 1 AGN) and more than two orders of magnitude in 15-$\mu$m flux (0.5 -- 60 mJy). 
The multi-band SEDs for these sources have been constructed by looking for counterparts in all the multi-wavelength catalogues
from far-UV to FIR available in the ELAIS-S1 field. Given for granted the optical R-band/spectroscopic association performed by La 
Franca et al. (2004), we have searched for associations in the {\em GALEX} Deep Imaging Survey (DIS), ESIS, $J, Ks$ and SWIRE catalogues.    
The cross-correlation radii between the 15-$\mu$m or optical positions and the different matching catalogues positions have
been chosen equal to the value above which spurious associations start dominating (i.e. Hook et al. 1998). 

The ultraviolet Galaxy Evolution Explorer ({\em GALEX}, Martin et al. 2005) Deep Imaging Survey (DIS) is observing $\sim$100 deg$^2$ in 
12 different areas of the sky, including ELAIS-S1 (Burgarella et al. 2005). Twelve ELAIS-S1 {\em GALEX} tiles are already available to 
the public as 
part of the second data release (GR2). The exposure times for these tiles vary from $\sim$3000 to $\sim$50000 sec. 
The photometric catalogues in the far-UV (FUV; 1530 \AA) and near-UV (NUV; 2310 \AA) have been cross-correlated with the optical
positions of the {\em ISOCAM} source counterparts, with a matching radius of 3 arcsec. Then all the possible UV counterparts have 
been visually inspected on the GALEX images, in order to reject spurious associations. We found 176 likely associations 
with the NUV catalogue, 169 of which are also in the FUV one.

The {\em ESO-Spitzer} Imaging Extragalactic Survey (ESIS; Berta et al. 2006) has covered the central 1.5 deg$^2$ of ELAIS-S1 
with WFI/2.2-m deep observations in the $B, V$ and $R$ bands, reaching 95\% completeness at $B, V\sim25$ and $R\sim$24.5.
The cross-correlation between the original optical positions and the ESIS positions (within 2 arcsec) have produced a good match for 90 sources
(those falling in the area covered by ESIS), thus providing additional optical ($B, V$) magnitudes for about 45\% of the 203
{\em ISOCAM} sources. In the area not covered by ESIS, $R$ magnitudes from the La Franca et al. (2004) catalogue and
$b_J$ magnitudes from APM survey ($\sim$90\% complete to $b_J = 20.5$; Maddox et al. 1990) have been used.

The central square degree of the ELAIS-S1 area (S1-5) was the target of NIR $J$ and $Ks$ imaging with NTT/SOFI, reaching 
$J\simeq$21 and $Ks\simeq$20 (Vega)\footnote{as part of the ESO Large Programme 170.A-0143, P.I. A. Cimatti} (Matute et al., in preparation).
The optical positions of the 15-$\mu$m source counterparts have been cross-correlated with the NIR catalogue ones, using a matching radius
of 2 arcsec. Thirty unique NIR counterparts have been found in the central square degree of ELAIS-S1, while additional
NIR data from the Two Micron All Sky Survey (2MASS; Jarrett et al. 2000) are available over the whole {\em ISO} area, providing
110 matches (10 of which are in common with the S1-$Ks$ ones). In total we found $Ks$-band magnitudes for 130 and $J$-band
magnitudes for 30 out of the 203 15-$\mu$m sources.   

The {\em Spitzer}/SWIRE observations in ELAIS-S1 cover the whole {\em ISO} region and a total sky area of $\sim$7 deg$^2$ in all 
the {\em IRAC} and {\em MIPS} bands, reaching 5$\sigma$ sensitivities of 4.1, 8.5, 48.2, 53.0, 256.0, 26$\times10^3$ and 166$\times10^3$ 
$\mu$Jy in the 3.6, 4.5, 5.8, 8.0, 24, 70 and 160 $\mu$m channels respectively (Lonsdale et al. 2004). The SWIRE data in ELAIS-S1 
have been released to the community in autumn 2005 (Data Release 3, DR3) through the query building GATOR at the NASA/Infrared Science 
Archive ({\tt http://irsa.ipac.caltech.edu/applications/Gator/}). 
However, to the purposes of this work, we had access to the SWIRE working catalogues, which contain all of the sources in 
the public catalogue, but reach deeper flux densities. 
Details about the SWIRE data reduction, that was carried out by the 
Spitzer Science Center and SWIRE team, can be found in the Data Release paper (Surace et al. 2005\footnote{also {\tt 
http://data.spitzer.caltech.edu/popular/swire/20050603$\_$enhanced$\_$v1/Documents/\\SWIRE2$\_$doc$\_$083105.pdf}}). 
For the {\em IRAC} fluxes and positions we used the band-merged working catalogue, consisting of {\em IRAC} and {\em MIPS}-24 $\mu$m fluxes 
associated with each other. While the public catalogue is a 3.6-$\mu$m based one, requiring detections at 3.6 and 4.5 $\mu$m above specific 
SNR thresholds (10 at 3.6 $\mu$m and 5 at 4.5 $\mu$m), the working catalogue contains all detections,
at any signal-to-noise ratios and even if a source is detected only in one {\em IRAC} band. 
For the 24-, 70- and 160-$\mu$m fluxes and positions, we used the 
single-band only catalogues, which contain all the sources detected only in the {\em MIPS} waveband in question. 
The {\em IRAC} positional accuracy is about 0.2$^{\prime \prime}$--0.4$^{\prime \prime}$ in all bands,
while the {\em MIPS} ones are $\sim$1$^{\prime \prime}$, 4$^{\prime \prime}$ and 8$^{\prime \prime}$ at 24, 70 and 160 $\mu$m 
respectively. Fixed aperture photometry is used for point-like objects, corrected for aperture losses (we have considered the
SWIRE ``aperture2'' (1.9 arcsec in radius) for both {\em IRAC} and {\em MIPS}), while ``ad-hoc'' aperture photometry (i.e. as large 
as needed to recover the whole flux of very extended sources)
has been performed by us directly on images for the extended sources, for which even the SExtractor's Kron fluxes
(those suggested for extended sources) are found to underestimate the total flux densities (up to a factor of $\sim$2 at 24 $\mu$m). 
For the match with the {\em IRAC} band-merged catalogue we have
considered the optical positions, searching for counterparts within a radius of 3 arcsec, while for match with the {\em MIPS}
catalogues we have considered the {\em ISOCAM} positions with searching radii of 5, 10 and 20 arcsec at 24, 70 and 160 $\mu$m
respectively. All but one of the 203 15-$\mu$m sources are detected in at least two {\em IRAC} bands (most in all 4 bands), while
200 are detected at 24 $\mu$m, 110 at 70 $\mu$m and 59 at 160 $\mu$m. The matches with 24-, 70- and 160-$\mu$m sources have been
visually inspected to be sure they were not mismatches. From the visual inspection of the {\em MIPS} images we have found
clear detections for 43 and 45 additional sources at 70- and 160-$\mu$m respectively. However, we have chosen not to derive
flux densities for these sources, considering in this work only those included in the SWIRE catalogues.

In Table 1 we report the multi-wavelength informations available for all the 203 15-$\mu$m sources. The 
source name from the Lari et al. (2001) catalogue is reported
in the first column, while the flux density (in $\mu$Jy) in all bands, from FUV to FIR, the redshift and the total infrared luminosity
(obtained by integrating the best-fitting SED in the range 8 -- 1000 $\mu$m) are reported in the 
following columns. In the last two columns the spectroscopic and SED-fitting classifications (see Section
\ref{class}) are presented. In case of no detection we have quoted a 3$\sigma$ upper limit, while $-$ means "no data coverage". 

\subsection{The Template Library}
\label{templ}
We have made use of all the available data, from FUV to FIR, to derive the SEDs of these 203 sources and construct the first 
highly complete observational sample of SEDs for MIR galaxies and AGN at intermediate $z$ to be used for statistical purposes (i.e. luminosity function). To interpret and classify the observed SEDs, 
we have performed a 
fit with several local template SEDs, representative of different classes of IR galaxies and AGN (including 3 Ellipticals of 
different ages, 1 lenticular, 7 Spirals, 3 Starbursts, 3 QSOs, 1 reddened QSO, Seyfert 1, 1.8 and 2 and 2 ULIRGs, containing 
both a starburst and an AGN component, in the wavelength range between 0.1 and 1000 $\mu$m (Polletta et al. 2007), appropriately redshifted to the rest frame. The full library of galaxy and AGN templates (excluding the three ellipticals since none
of our sources could be fitted by any of them) is shown in Fig. \ref{fig_templ}. 
The elliptical, spiral and starburst templates were generated with the
GRASIL code (Silva et al. 1998). The spiral templates range from early to late
types (S0-Sdm), the starburst templates correspond to the SEDs of NGC 6090,
M 82 and Arp 220. Templates of moderately luminous AGN, representing Seyfert
1.8 and Seyfert 2 galaxies, were obtained by combining models and data from
a large sample of Seyfert galaxies. The AGN templates include three
templates representing optically-selected QSOs with different values of
IR/optical flux ratios (QSO, QSO$\_$high, and QSO$\_$low) and one reddened QSO
(red$\_$QSO). The composite (AGN+SB) templates are empirical templates created to
fit the SEDs of the heavily obscured BAL QSO Mrk~231 (Berta 2005) and 
the Seyfert 2 galaxy IRAS~19254$-$7245 South (Berta et al. 2003). These objects contain a
powerful starburst component, mainly responsible for their FIR emission,
and an AGN component that contributes to the MIR (Farrah et al. 2003).
With respect to existing template libraries derived from empirical
SEDs (Coleman et al. 1980) or from models (Bruzual 2003; Fioc et al. 1997; Silva
et al. 1998; Devriendt 1999), this library has a broader wavelength coverage and variety of
spectral types. Examples of application of this library to various types of
SWIRE sources can be found
in Lonsdale et al. (2004), Franceschini et al. (2005), Hatziminaoglou et al. (2005),
Jarrett et al. (2006), Polletta et al. (2006), Weedman et al. (2006), Tajer et al. (2007),
Polletta et al. (2007).

To avoid biasing the results in favour of a small subset of templates, we adopt the full library of Polletta et al. (2007) to fit the SEDs. Although some templates are obtained from modelling the SEDs of Ultra-Luminous objects (ULIGs, $L_{IR} > 10^{12} L_{\odot}$: e.g. Mrk 231, I19254 and Arp 220), they are kept in the library since they well represent a large fraction of the Spitzer population at several flux and luminosity levels (see e.g. Franceschini et al. 2005, Polletta et al. 2006, 2007, Alonso-Herrero et al. 2006).

%

\subsection{The Fitting Method}
\label{fitting}
The SEDs of our sources are fitted using an automated fitting routine contained in the {\em Le 
PHARE} software (by Arnouts \& Ilbert, available at {\tt http://www.lam.oamp.fr/people/\\
arnouts/LE$\_$PHARE.html}),
offering the possibility of using any template library and filters set. {\em Le PHARE} (PHotometric Analysis 
for Redshift Estimations; Ilbert et al. 2006) is a publicly available set of fortran programs aiming at 
computing photometric redshifts through the best-fitting SED analysis. 
%
%
The program is based on a simple $\chi^2$ fitting method between the expected magnitudes
and an observed photometric catalogue.

Since we already know the spectroscopic redshift of our galaxies, we have
fixed $z$ and have used {\em Le PHARE} only to obtain the best-fitting SED through a comparison
with the template SEDs. 
No extinction was added to the templates, since some of them are already intrinsically extinct.
Note that by using a spectroscopically complete sample we are able to avoid all the uncertainties due to 
the photometric redshift measurements, which greatly increase the degeneracy in the SED-fitting procedure. 
Moreover, for the first time we can compare the results from optical spectroscopy with the broad-band SED shape 
of a statistically significant sample of MIR galaxies and AGN.

We make the basic assumption that the SED shapes seen at low redshifts can also well represent the higher redshift objects. At the end of our procedure we would be able to verify this assumption.
The template library used to fit our data contains a finite number of SEDs (21), representative of given 
classes of local infrared objects, which do not vary with continuity from one class to another (there are large gaps in the parameter space). Therefore, the quality of the fit depends not only on the photometric errors, but also on the template SED uncertainties. For this reason, in our fitting procedure, in addition to the photometric errors on data, we need to take into account also the uncertainties due to the template SEDs ``discretization''. To estimate and consider the total uncertainty on both the photometric data and the template SEDs, we have proceeded as follows.
First, we have increased all the formal catalogue errors (probably too small) by a reasonable given amount, going from 5\% in the UV,
and optical bands up to $\sim$15\% in the {\em MIPS} bands.
We have run {\em Le Phare} on all the sources, obtaining a first "estimate" of the best-fitting SED classification.
We have then computed the distributions of the ($S_{object}-S_{template}$)$_{band}/(\sigma_{object})_{band}$ values
in each of the considered photometric band (where $S_{object}$ and $\sigma_{object}$ are the flux density and the relative error
of the source, and $S_{template}$ the flux of the template in the considered band), iteratively increasing the photometric 
errors until we have obtained a Gaussian 
distribution with $\sigma \simeq 1$. This corresponds to reduced $\chi^2$ distributions peaked around 1
(as should be in case of good fit). The enlarged relative errors ($\sigma_{object}/S_{object}$) obtained in this way 
go, on average, from $\sim$12--15\% in the optical/NIR/{\em IRAC} bands up to $\sim$20\% in the {\em MIPS bands}.
With the new (significantly larger) photometric uncertainties, which now take into account also the additional uncertainties induced by the template SED discretization, we have run {\em Le PHARE} on our 
sources for the second time, obtaining what we have taken as the final SED-fitting results.  

\subsection{The SED-Fitting Results}
\label{fit}
\begin{figure}
\plotone{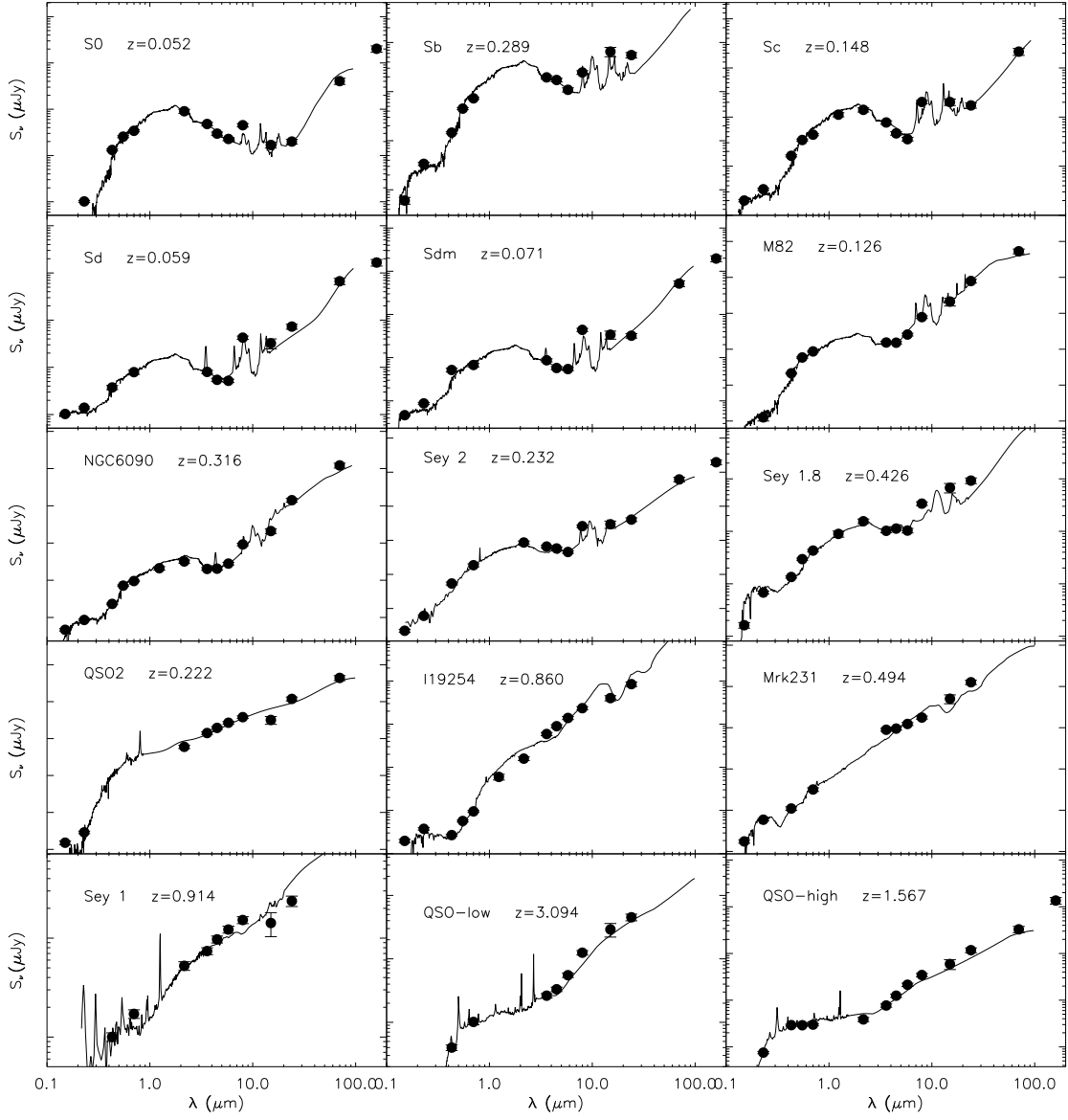}
\caption{Example of SED fits for our sample: the data (filled circles with error-bars), plotted 
as flux ($S_{\nu}$ in $\mu$Jy) versus observed wavelength, are over-imposed to the corresponding 
best-fitting template SEDs (solid line), opportunely redshifted at the source's $z$. The plotted uncertainties 
are those reported in the various catalogues and not those used for the fits.
}
\label{fig_fit}
\end{figure}

In figure \ref{fig_fit} examples of data overlying the best-fitting template SEDs are shown for most of the 
templates reproducing our data (the original library contains 21 SEDs, but none of our sources could be fitted by any of 
the 3 elliptical galaxy templates). We find that 83 sources (41\%) are reproduced
by a galaxy template SED (4 S0, 3 Sa, 16 Sb, 27 Sc, 18 Sd, 15 Sdm), 13 (6.5\%) with a starburst template SED
(1 Arp~220, 5 M~82, 7 NGC~6090), 72 (35\%) with a type 2 AGN template SED (52 Seyfert 2, 15 Seyfert 1.8, 5 red QSO), 11 (5.5\%) with a
ULIG composite SED (9 Markarian~231, 2 I19254) and 24 (12\%) with a type 1 AGN SED (3 Seyfert 1, 6 QSO, 13 high IR/optical 
flux ratio QSO, 2 low IR/optical flux ratio QSO). Therefore, with our broad-band SED-fitting method we find AGN activity in 52.5\% of our MIR selected sample.

\begin{figure}
\plotone{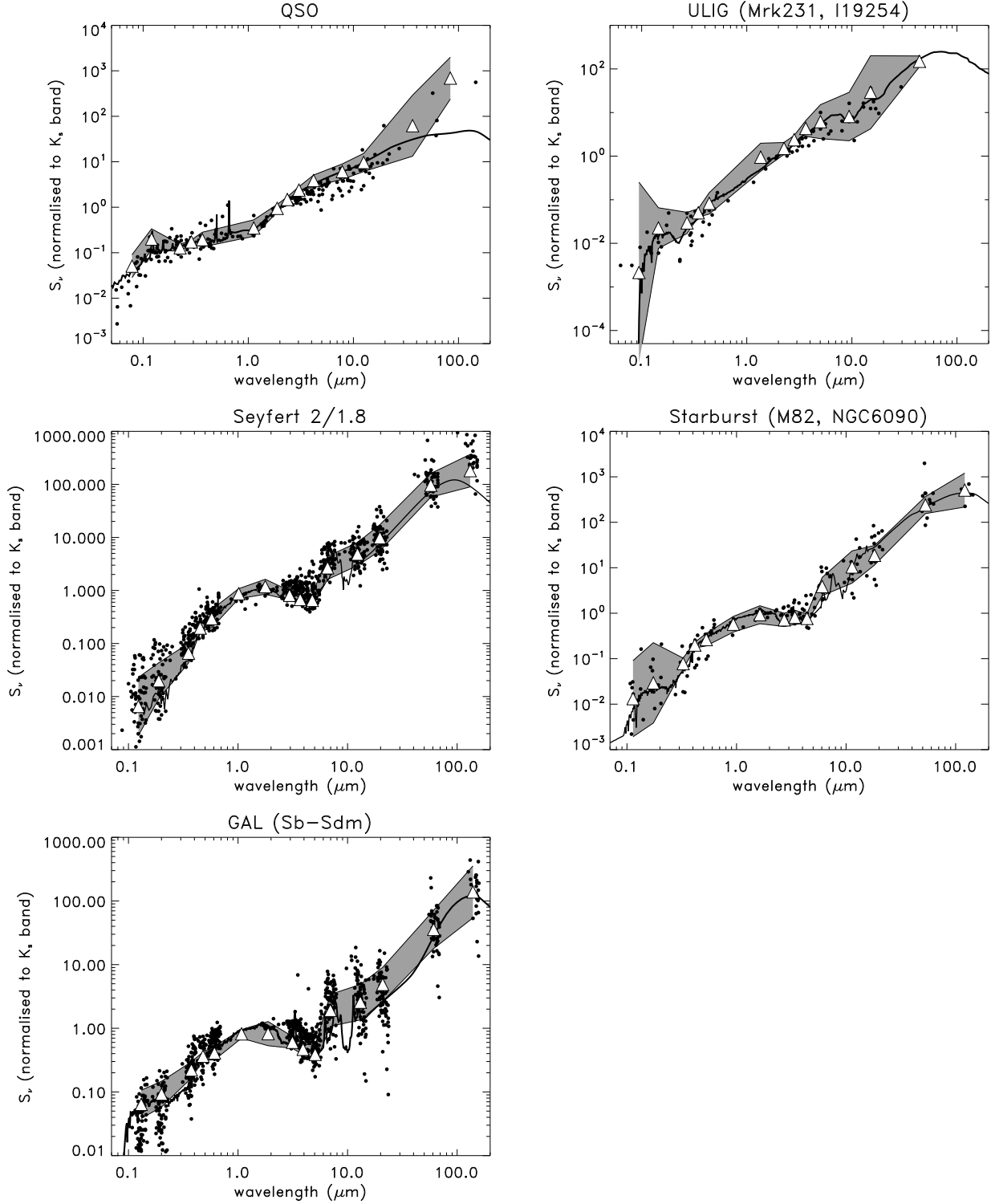}
\caption{Rest-frame SEDs (dots) of 15-$\mu$m sources belonging to the different broad SED classes
(AGN1, ULIG, AGN2, Starburst and Spiral galaxy) compared to the most frequent best-fit template SED of
that class, all normalised to the $Ks$-band flux density. The white triangles represent the 
average values that the templates should have in the different rest-frame bands to better reproduce
our sources, while the grey-shaded areas show their relative uncertainty region corresponding to $\pm 1\sigma$.  
}
\label{fig_uncert}
\end{figure}

We note that the distributions of the final $log_{10}(S_{object}/S_{template})$ values in some 
bands and for some SED classes are centred on values slightly different from 0. These offsets, which are different
in the different bands and for the different best-fitting SED classes (i.e. for the QSO class the larger offsets are observed
at 70 and 160 $\mu$m, while for the galaxy class the larger offsets are observed in the B, J, K, 15-$\mu$m and 24-$\mu$m bands; see Fig. 
\ref{fig_uncert}), are interpreted as the average values we should add to the template SEDs to better reproduce the data ensemble
of MIR sources. In Fig. \ref{fig_uncert} we show the rest-frame SEDs (dots) of 15-$\mu$m sources belonging to the different ``broad'' 
SED classes (AGN1, ULIG, AGN2, Starburst and Spiral galaxy), compared to the most frequent best-fit template SED of
that class (i.e. ``QSO'' for AGN1, ``Seyfert 2'' for AGN2, ``Mrk 231'' for ULIG, ``NGC 6090'' for Starbust, ``Sc'' for
Spiral), all normalised 
to the $Ks$-band flux density. The white triangles represent the average values that the templates should have in the different rest-frame 
bands to better reproduce the bulk of our sources, while the grey-shaded areas represent their relative uncertainty region corresponding to $\pm 1\sigma$ of the expected values. From this check we can conclude that, on average, the template SEDs of the Polletta et al. (2007) library modelled on local galaxies and AGN are able to reproduce also most of the MIR selected sources at intermediate redshifts, though some of the local templates should be slightly modified in some wave-bands to get a better agreement with the bulk of the observed data.  
     
In order to estimate the reliability of the SED classification, we compared the classification
and $\chi^2$ obtained by the two best solutions of each fit: the primary (corresponding to the minimum $\chi^2$ value, 
$\chi^2_{best}$) and the secondary solutions ($\chi^2_{2nd}$). We find that the primary solutions of our photometric 
analysis are quite stable within uncertainties. In fact, for the majority of our sources (92\%) the $\chi^2_{best}$ values
are within the 90\% probability threshold ($\chi^2_{90\%}=21.07$ for the case of 14 degrees of freedom).
In most cases the secondary solution belongs to the same ``broad'' SED class of the primary (i.e. AGN or galaxy, containing all the type 1's, 
type 2's and ULIG templates and all the galaxy and starburst templates, respectively) and only for
23 objects (11\%) it belongs to a different class, with a $\chi^2_{2nd}$ lower than the $\chi^2_{90\%}$ threshold. In all the other
cases, if the primary and secondary solutions correspond to different SED classes, the secondary solutions have a low
probability of being acceptable ($\chi^2_{2nd} > \chi^2_{90\%}$). The 16 sources (8\%) with a statistically bad primary solution 
($\chi^2_{best}$ $> \chi^2_{90\%}$) have, in a few cases, some evident photometric errors in one or more bands (mainly in the
$J, K^{'}$ ones). In all the remaining cases, the bad fit is due to an observed SED flatter than the best-fitting template in the MIR domain, 
thus implying the need of a template SED (missing in our library) intermediate between the power-law and the Seyfert 2/1.8 ones,
where the AGN is not dominant as in the type 1 template case (power-law), but its contribution is higher than in the Seyfert 2 
template case. 


\section{Spectroscopic versus SED Classification}
\label{class}
We have compared the results of our SED-fitting procedure with the results obtained from the optical spectroscopy. 
In Table \ref{tab_class} we show for each spectroscopic class the number of sources which are best fitted by the various SED templates and vice-versa. The reported percentages are the fractions
of sources of a given spectroscopic classification fitted by the various template SEDs. 
The spectroscopic type 1 (broad line) AGN are all fitted by AGN template SEDs, and therefore the agreement between the two 
classifications for type 1 AGN is very good (the only three objects fitted by a type 2 AGN template are indeed fitted by a Seyfert 
1.8 template). 
However, the agreement is not equally good for the spectroscopic type 2 AGN ($+$ LINERs) and galaxies: 56\% of the AGN2$+$LINER 
sample is fitted by a type 
2 AGN template and another 19\% is fitted by type 1 AGN or composite (AGN+starburst) templates, but $\sim$25\% do show a galaxy SED. 
Also for the majority (63\%) of the spectroscopic normal galaxies the two classifications agree (another 3\% is fitted by a starburst
template), but there is a significant 
fraction (34\%) fitted by AGN templates (mostly type 2's). For the spectroscopic starbursts the fraction of
sources classified as AGN by the SED fitting is even higher, 53\%. The small number of spectroscopically unclassified sources is 
fitted for the 40\% by AGN and for the 60\% by galaxy templates.
Therefore, we can conclude that the SED-fitting technique applied to our MIR selected sample is able to identify AGN activity 
in $\sim$40\% of sources spectroscopically unclassified, in $\sim$34\% of sources spectroscopically classified as galaxies, 
in $\sim$53\% of spectroscopic starbursts, in 75\% of type 2 AGN or LINERs, and in 100\% of type 1 AGN. 
The main result of this comparison is that, although for 
many sources the spectroscopic classification is in agreement with the classification resulting from the SED-fitting method, 
the broad-band SED analysis is able to find AGN activity in a higher fraction of MIR sources ($\sim$53\%) than the optical 
line diagnostic techniques ($\sim$29\%). On the other hand, a small number (14\%) of the spectroscopically
classified AGN are classified as galaxies by the SED fitting.

For comparison, we cite the recent results of Polletta et al. (2007), who, using a sample of X-ray selected AGN with 
available spectroscopic classification and a similar fitting method as applied to our sample, find that 82\% of the X-ray selected AGNs are classified as AGN by the SED-based classification method, compared to 78\% by the spectroscopic classification. 
However, the SED and the spectroscopic classification agree only in 53\% of the cases: the SED method and the spectroscopic classification for X-ray sources are consistent for all type 1 AGN, but for only 29\% of type 2 AGN and 33\% of AGN with galaxy spectra.

The more evident change between the spectroscopic and SED classifications in our sample is due to the fact that several sources 
optically classified as normal or starburst galaxies do show a Seyfert 2/Seyfert 1.8-like SED, or possibly a composite 
(starburst+AGN) SED. Therefore, the total fraction of type 2 AGN increases from 15.5\% (spectroscopic) to 36\% (SED-fitting), 
while that of galaxies powered by star-formation (normal+starburst) decreases from 70\% (spectroscopic) to 48\% (SED-fitting). 
In addition, we have 4\% of composite (AGN+starburst) SEDs that were not considered as spectroscopic class. 

\begin{figure}
\plotone{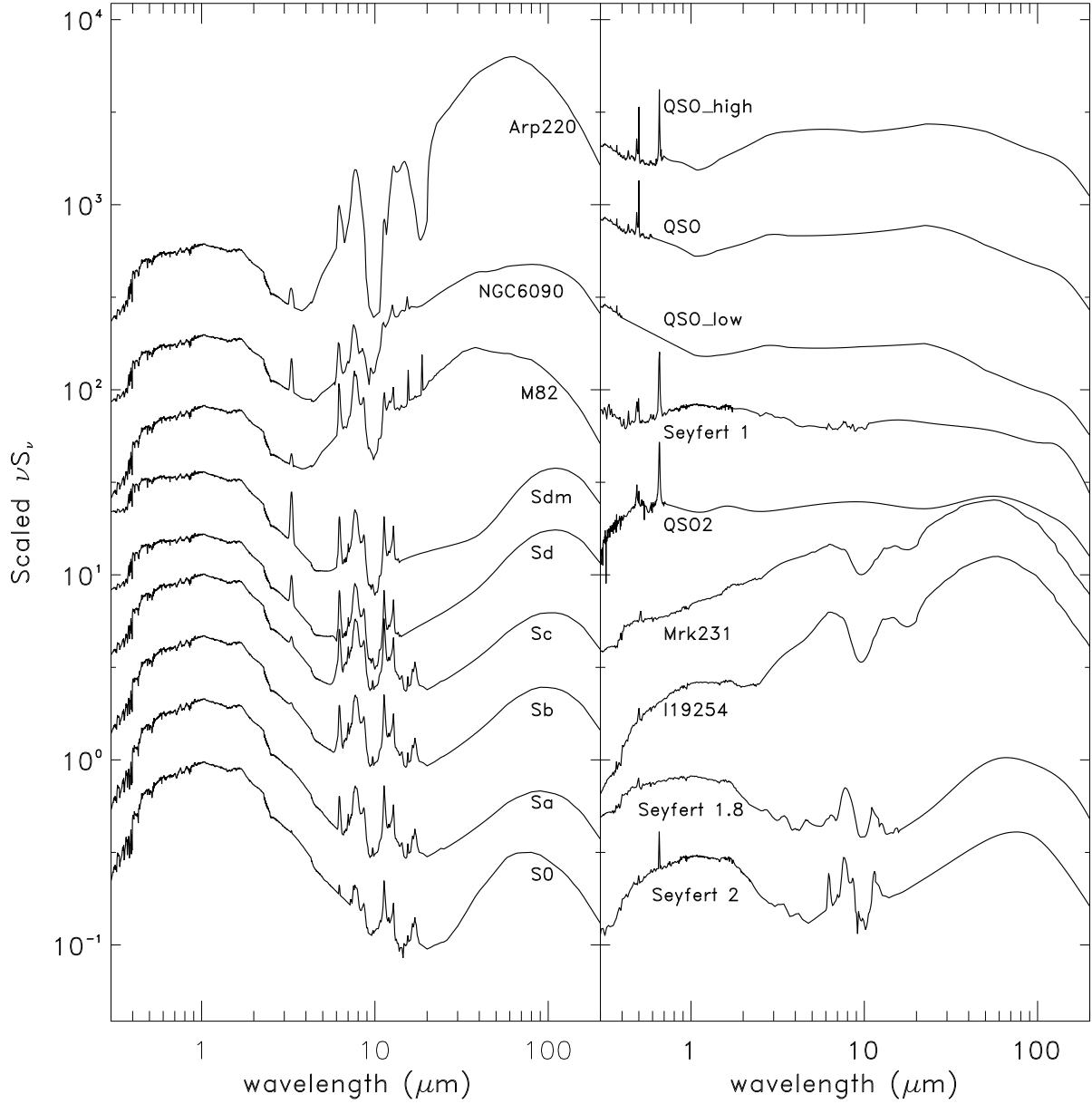}
\caption{Library of template SEDs from Polletta et al. (2007). The SEDs are plotted
in arbitrarily scaled luminosity $\nu S_\nu$ versus wavelength.   
}
\label{fig_templ}
\end{figure}

As a partial explanation of these discrepancies, we note that many of the sources spectroscopically 
classified as normal galaxies do not have enough lines in their optical spectra to be properly classified according to the 
standard line ratios diagnostics (i.e. Tresse et al. 1996).
In fact, many of them have only $H\alpha$ and/or $[OII]$ in emission and therefore the spectroscopic classification 
might be misleading. However, in all the cases where the classification changed from spectroscopy to SED-fitting, the 
spectra do either show no AGN emission lines (e.g. [OIII]) at all, or lower than those due to star-formation, 
making them clearly classified as galaxy (normal or starburst) through the diagnostics diagrams.
In any case, the SED-fitting procedure seems to find more AGN than the optical spectral lines
classification. One possible reason is that in most of these objects the AGN dominates the energetic output
only in the MIR.
Indeed, most of the sources in our sample with galaxy spectroscopic classification and SED reproduced by a Seyfert 2/1.8 template
could be fitted by a normal galaxy in the optical/NIR part of their spectra, but not in the MIR (in particular 
in the 3--8 $\mu$m range), where the data are too flat to be reproduced by a normal/starburst galaxy SED and 
therefore require the presence of a low-luminosity AGN. The AGN shows up just in the range where the host galaxy 
SED has a minimum, due to the junction between the stellar photospheric emission (dominating the optical/NIR) and the dust component 
(starting dominating at $\lambda>5\mu$m; see e.g. the late-type spirals and starburst templates in Fig. \ref{fig_templ}), 
which is also the range where the hot dust heated by an AGN is expected to start increasing towards the peak.   
In case an AGN component is present with a NIR luminosity similar to that of the host
galaxy, emission from hot dust heated by the AGN contributes to the NIR/MIR
filling up the dip observed in star-forming galaxies and producing a flat
NIR/MIR spectrum (see e.g. the Seyfert templates in Figure \ref{fig_templ}).
This is consistent with the results about X-ray selected obscured AGN, which are known to be ``elusive'' in their optical spectra 
due to host galaxy dilution or heavy dust and gas obscuration or both (i.e. see Fiore et al. 2003; Maiolino et al. 2006; Tajer et al. 2007; Caccianiga et al. 2007; Civano et al. 2007; Cocchia et al. 2007).
Unfortunately, a check with the X-ray luminosity for our ``elusive'' AGN is inconclusive for our purposes, since the XMM observations 
cover only a small portion of the field (15\%).

In order to visualise what discussed above and look at the average properties (in terms of redshift and rest-frame luminosity) of the 
different SED classes, in Fig. \ref{fig_histo} we show the redshift (in logarithmic scale) and the log(L$_{15 \mu m}$) distributions for 
the different SED classes to which our sources belong. 
We notice a clear trend of increasing 15-$\mu$m luminosity 
(and $z$) from early-type (S0, Sa) to later-type (Sd, starburst) SEDs. Type-2 AGN are spread over almost the entire range 
of L$_{15 \mu m}$ ($\sim$10$^9$ -- $10^{12}$ L$_{\odot}$, except two $z > 2$ objects at L$_{15 \mu m} > 10^{13}$ L$_{\odot}$)
and over a large redshift range (0.05$<z<$3), while type 1's occupy the region of higher luminosities ($L_{15 \mu m} > 10^{10}$ 
L$_{\odot}$) and redshifts (most of them are at $0.5 < z < 3$).
The different spectroscopic classifications have been
highlighted by filling the relative distributions with different patterns, as shown in the legend within the plot.

Although very uncertain, since the templates often do not fit well the 70 and 160 $\mu$m data, the total IR luminosities ($L_{IR}$) have 
been computed by integrating the best-fitting template
SEDs between 8 and 1000 $\mu$m. Most of the starburst galaxies and some of the type 2 AGN are in the ULIG 
luminosity range ($L_{IR} > 10^{12}$ L$_{\odot}$), with the remainder in these two classes in the LIG range ($10^{11} < L_{IR} < 10^{12}$ L$_{\odot}$). 
The majority of normal galaxies have $L_{IR} < 10^{11}$ L$_{\odot}$, with just some in the LIG range. All the type 1 AGN are in the ULIG range, with few even in the Hyper-LIG (HyLIG) range ($L_{IR} > 10^{13}$ L$_{\odot}$), where all the composite AGN are indeed.

\section{Discussion and Conclusions}
\label{discuss}

We have derived the broad-band (from FUV to FIR) SEDs for the largest available highly complete (72\%) sample of 
intermediate-$z$ MIR selected galaxies and AGN with spectroscopic identification and redshift. 
The majority of these objects are at $0.1 < z < 1.3$ and in the flux density range where strong evolution is observed in the
counts of MIR sources. Their broad-band SED characterisation is therefore fundamental to understand
the nature of sources responsible for the observed evolution. 
Based on the SED-fitting technique we have classified the MIR sources, identifying 
AGN signatures in about 53\% of them. This fraction is significantly higher than that derived 
from optical spectroscopy ($\sim$29\%) and is due principally to the identification of 
type 2 AGN activity in objects spectroscopically classified as galaxies.   
This might be caused, at least in part, by host galaxy dilution in the optical, similarly to what is observed for
obscured AGN selected in X-ray. It is likely that in most of our objects where the AGN
luminosity in the optical band is fainter than that of the host galaxy, the AGN dominates the energetic output only
in the MIR, showing up just in the range where the host galaxy SED has a minimum,
while the hot dust heated by the AGN is expected to start peaking.

We note that the fraction of MIR sources with an ``elusive'' AGN derived in this work is in agreement with recent
results from MIR spectroscopy (with {\em Spitzer-IRS}) of local star-forming galaxies from the {\em Spitzer}-SINGS
 sample
by Smith et al. (2007), who find that a high fraction ($\sim$50\%) of local galaxies do harbour low-luminosity AGN (LINER or Seyfert types).
These low-luminosity AGN, contrarily to the luminous ones which completely destroy PAH grains, are able to excite 
the PAHs and to modify the emitting grain distribution, thus providing unusual PAH spectra with very weak 
or absent 6.2 $\mu$m, 7.7 $\mu$m and 8.6 $\mu$m bands, suppressed in strength by up to a factor of 10. 
An indication of the presence of such weak AGN systems, which are found to shift power from the short to the
long wavelength PAHs, can be revealed, for example, by the strength ratio between the PAH features 
at 7.7 $\mu$m and at 11.3 $\mu$m (Smith et al. 2007). Therefore, the definitive confirmation of the presence of 
weak AGN systems in our intermediate-$z$ MIR dusty galaxies, at present, could be obtained only through {\em IRS} 
spectroscopy, allowing to test with line ratios their SED shape indications. 

\begin{figure}
\plotone{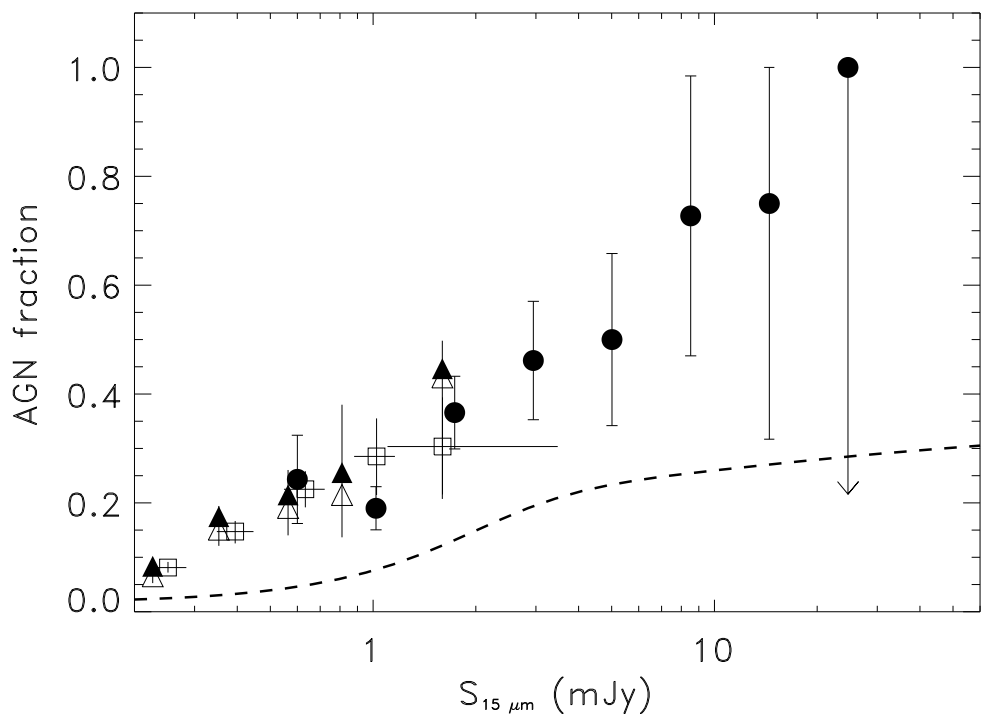}
\caption{Fraction of AGN over total extragalactic sources as a function of the 15-$\mu$m flux density,
as derived from the spectroscopic classification (dashed line; Matute et al. 2006) and from the
SED-fitting performed in this work (filled circles with error-bars). For comparison, the results of Brand et al. (2006) and Treister et al. (2006)
have been plotted as open squares and triangles (open: non corrected for AGN not detected in X-rays; filled: corrected for AGN not detected in X-rays), respectively.
}
\label{fig_frac}
\end{figure}

Given the new percentages of AGN, we have updated the relative fractions of AGN and galaxies
contributing to the observed 15-$\mu$m source counts. 
Although the results of our work might still be affected, at some level, by photometric errors in the 
data and by some degrees of degeneracy in the template SEDs, 
the analysis of the photometric errors and the agreement between the primary and secondary solutions in the
vast majority of the cases make us confident that the majority of the SED-classified AGN are reliable.
For coherence with the previous classification and
source counts (see La Franca et al. 2004 and considerations therein) and to avoid biasing towards AGN, we have treated all the sources without 
spectroscopic redshift as galaxies. In fact, a spectroscopic sample 72\% complete can be biased towards AGN since the large majority of the spectroscopically unidentified objects have 15-$\mu$m flux densities fainter than 1 mJy. Galaxies dominate over AGN in this range of flux and it is therefore likely that most of the unidentified objects are galaxies. 
On the other hand, since we know that optical spectroscopy can miss AGN signatures, we can 
take the fraction of spectroscopic AGN (used for deriving the AGN LF at 
15 $\mu$m by Matute et al. 2006) as a lower limit. 
In Fig. \ref{fig_frac} we show the fractions of AGN as function of 
15-$\mu$m flux obtained by considering the spectroscopic classification 
(dashed line; from Matute et al. 2006) and the SED classification (filled circles with error-bars; this work). 
Other recent works (Brand et al. 2006; Treister et al. 2006), though based on different diagnostics 
(e.g. the 24 $\mu$m to 8 $\mu$m flux ratio or X-ray detection), estimate lower
limits to the AGN contribution in MIR surveys significantly higher than those predicted 
by Matute et al. (2006) and in very good agreement with the fractions derived in this paper
(see Fig. \ref{fig_frac}, where the Brand et al. 2006 and Treister et al. 2006 results are reported
for comparison, with the 24-$\mu$m flux densities converted to 15 $\mu$m by using the opportune 24/15 $\mu$m 
ratio as function of flux as computed by Gruppioni et al. 2005).
Such results might have a significant impact on the interpretation of galaxy and AGN evolution and on the 
physics of the MIR selected objects, so far often erroneously assumed to be all starburst galaxies 
in many models. 
In fact, all the existing models of galaxy and AGN evolution in the IR should be revised by taking 
into account the higher fraction of AGN dominating in the MIR wave-range. 

\begin{figure}
\epsscale{1.02}
\plotone{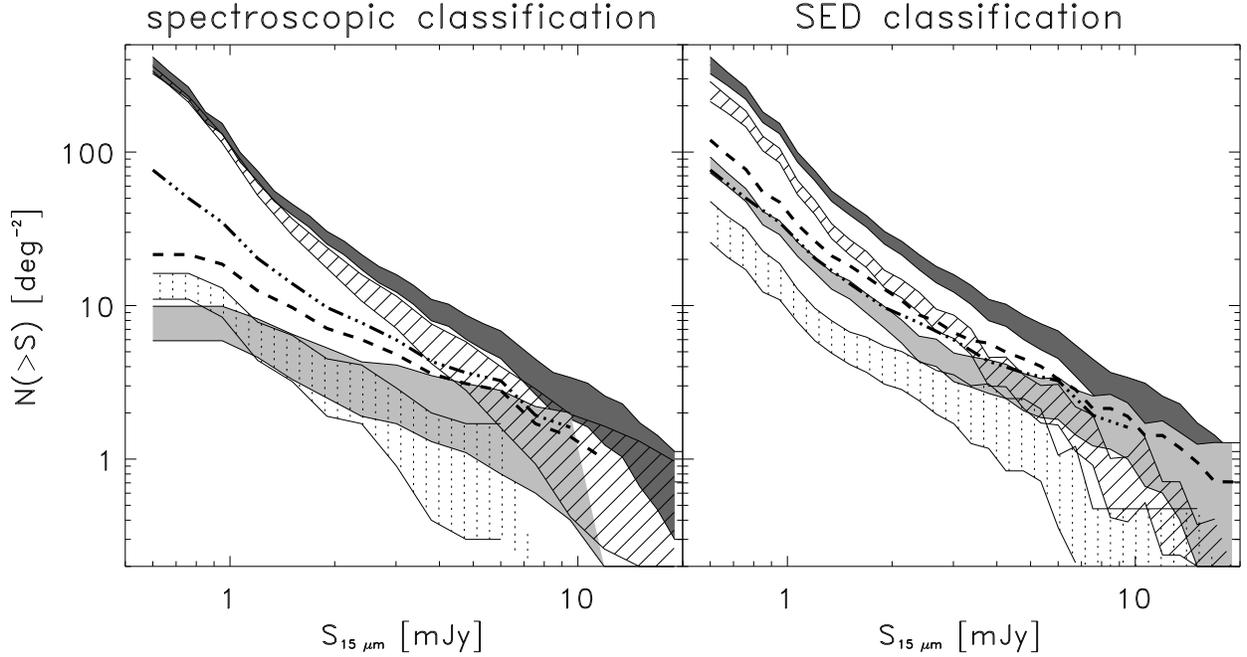}
\caption{Observed extragalactic source counts at 15 $\mu$m in ELAIS-S1: total counts with uncertainties 
(dark-grey shaded area); type 2 AGN contribution (light-grey shaded area); type 1 AGN contribution (dotted area);
galaxy contribution (hatched area). The total (type 1 + type 2) AGN contribution is shown by the dashed line,
while the dot-dot-dashed line represents the total AGN counts at 15 $\mu$m updated by La Franca et al. (2007)
on the basis of hard X-ray observations. 
In the $left$ panel the AGN/galaxy contributions are derived from the spectroscopic classification (La Franca et al. 2004),
while in the $right$ panel they come from the SED-fitting work described in this paper. All the optically
unidentified sources have been conservatively considered as galaxies in both plots (see Section \ref{discuss}).
}
\label{fig_count}
\end{figure}

By comparing the AGN fraction derived by our SED-fitting 
analysis to the ``lower-limit'' coming from the spectroscopic classification, we can determine how 
the relative source counts of galaxies and AGN (and consequently the relative evolutionary models) 
should change according to the new values. 
To this purpose, in Fig. \ref{fig_count} we have plotted the extragalactic source counts at 
15 $\mu$m in the S1 field (Gruppioni et al. 2002) with the relative contribution of galaxies 
and AGN computed according to the previous ($left$) and new ($right$) determination of AGN fractions.
The difference between the two is small at bright fluxes, but the SED fitting method finds more and more AGN going to
fainter fluxes. The galaxy population now dominates below $\sim$3 mJy (previously 6--7 mJy), although now the AGN counts are only a factor of 2 lower even at
the lower flux densities.
For comparison, we have also shown the results of a recent work by La Franca et al. (2007), who updated the
AGN source counts at 15 $\mu$m on the basis of X-ray band ({\em XMM}) observations  (Puccetti et al. 2006)
on the central 0.6 deg$^2$ of ELAIS-S1.
By classifying as AGN all the MIR sources with an unabsorbed 2-10 keV X-ray luminosity $> 10^{42}$ erg s$^{-1}$, 
the authors find that at least 13\% of the previously classified galaxies on optical basis (by La Franca et al. 2004)
do harbour an AGN, therefore concluding that $> 24\%$ of the 15-$\mu$m sources with flux density $> 0.6$ mJy
are AGN. The AGN source counts at 15 $\mu$m updated by La Franca et al. (2007) and plotted in Fig. \ref{fig_count}
as dot-dot-dashed line are indeed significantly higher that those based on optical classification, and in better
agreement with (although somewhat lower than) the AGN counts derived in this work. The higher efficiency of the SED-fitting technique
with respect to X-ray luminosities in detecting AGN activity, in this specific case, might be partially 
due to the fact that {\em XMM} observations in S1 are not deep enough (S(2--10keV) = $2\times10^{-15}$ erg cm$^{-2}$ s$^{-1}$)
to detect (and thus allow conclusive results for) all (most of) our sources. Moreover, it is known that even the deepest X-ray surveys fail in detecting highly obscured 
(i.e. Compton thick: $N_H > 10^{24} cm^{-2}$) AGN, while sensitive measurements in the IR range, and in particular
SED studies in the NIR/MIR range, provide a unique and efficient opportunity to recover the fraction of obscured 
or ``elusive'' AGN not identified in X-ray surveys (Alonso-Herrero et al. 2006; Fiore et al. 2008; Martinez-Sansigre et al. 2007; Daddi et al. 2007). 

We have compared our results to those of Hickox et al. (2007), though the sample selected by those authors is an {\em IRAC} selected one fulfilling also additional selection criteria (i.e. detection in all the {\em IRAC} bands and in the $R$ band, {\em IRAC} colours falling in the Stern et al. 2005 AGN selection region, spec- or photo-$z>$0.7), finding that according to the Hickox et al. criteria just a few of our AGN can be considered ``obscured''. In fact, very few of our AGN do fall above the $R-[4.5]$=6.1 separation boundary considered by Hickox et al. (2007) to divide ``obscured'' from ``unobscured'' AGN. All our type 1 and most of type 2 are below that boundary, while most of the composite AGN are above. This would imply that most of our composite AGN are ``obscured'' and all our type 1 and most of our type 2 AGN are ``unobscured''.
The cumulative redshift distribution of our $z > 0.7$ AGN is in good agreement with that derived by Hickox et al. (2007), up to $z=1.2-1.3$. 
Above $z=1.3$ there are just few sources in our sample, due to our relatively bright 15-$\mu$m and $R$-band selection, therefore we can consider our AGN (in particular type 2) sample complete only up to $z=1.3$.  

The results of this work will be very useful for updating all the models aimed at interpreting the deep 
infrared survey data and, in particular, for constraining the nature and the role of dust-obscured 
systems in the intermediate/high-redshift Universe. The main changes to the actual evolutionary models for
IR galaxies and AGN would consist in better defined evolutionary source classes (and SEDs) and in updated
evolutions for the different classes. In particular, referring to the specific case of the Pozzi et al. (2004)
and Matute et al. (2006) models, which are based on the same sample considered here, although star-forming
galaxies are still the dominant class at low flux densities, their evolution should be lower than that derived by 
Pozzi et al. (2004). Similarly, the AGN (mainly type 2's) should evolve more rapidly (and probably
both in luminosity and density) than derived by Matute et al. (2006), similarly to star-forming galaxies.
In fact, the higher fraction of AGN in infrared surveys than found through optical spectroscopy  is a result common to
different works (i.e. Brand et al. 2006; Treister et al. 2006) and applies mainly to type 2 AGN. Type 1 AGN are in fact
easier to identify in optical, due to their broad lines, thus most (all) of them are classified from their optical spectra.
Type 2 AGN are more ``elusive'' and difficult to reveal in the optical. Therefore, the higher fraction of IR AGN is mainly 
due to the unveiling of type 2 AGN previously erroneously classified as galaxies. For this reason, the starburst galaxies, to whom 
all the evolution observed in the MIR was commonly attributed, are probably in a smaller number than formerly believed and part of their
evolution should be ascribed to type 2 AGN. 
This is also evident from the right panel of Fig. \ref{fig_count}, where the revised AGN source counts increase towards fainter fluxes with a slope 
similar to (though slightly lower than) that of galaxies, while, according with the results based on optical spectroscopy (left panel), the AGN
were rapidly converging towards the lower flux densities. 

\begin{figure}
\epsscale{0.9}
\plotone{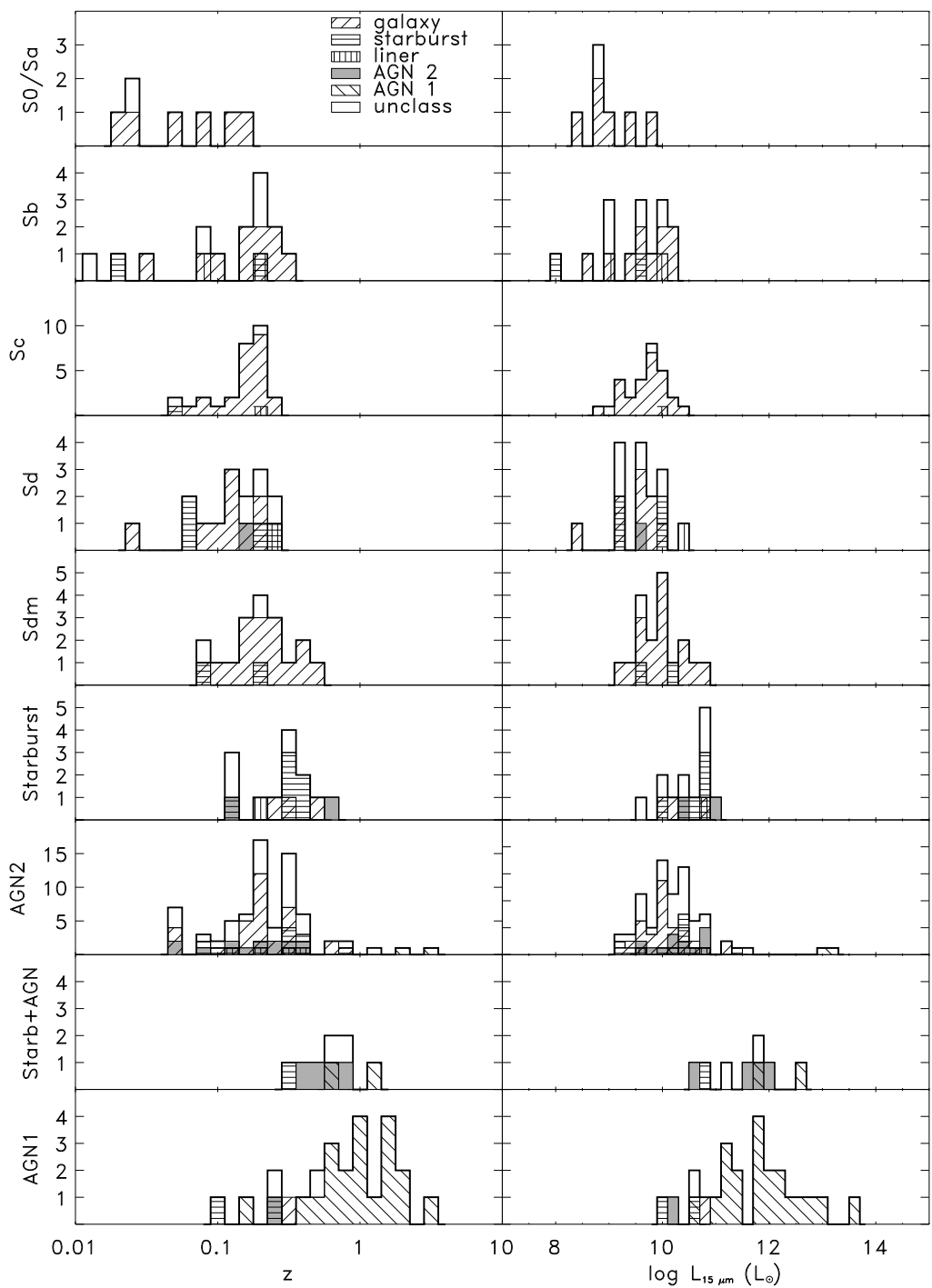}
\vspace{-1cm}
\caption{Logarithmic distributions of $z$ ($left$) and L$_{15 \mu m}$ luminosity ($right$) 
for the different SED classes reproducing our data. The luminosities have been derived by 
considering for each source its best-fitting template SED.
Within each distribution the different spectroscopic classes are highlighted, as explained in the legend.
}
\label{fig_histo}
\end{figure}

Finally, instead of considering non-evolving normal galaxies
(and therefore ascribing all the evolution to starburst galaxies), it would be more appropriate to consider galaxy SEDs evolving with $z$ (and/or $L$)
from early-type to late-type (and to starburst), as suggested by the results shown in Fig. \ref{fig_histo}, and eventually different evolutions for different luminosity intervals.
The revision of evolutionary models for IR sources based on the results presented here is 
beyond the aims of this work and will be treated in a forthcoming paper (Gruppioni et al., in preparation).

To summarise, in this paper through the broad-band SED analysis of the largest available
highly (72\%) complete spectroscopic sample of MIR selected galaxies and AGN at intermediate $z$, we have:
\begin{itemize}
\item verified the assumption that local template SEDs are able to reproduce also most of the MIR-selected galaxies and AGN at $0.1 < z < 1.3$;
\item found AGN activity in a significantly higher fraction of sources ($\sim$53\%) than derived from optical spectroscopy ($\sim$29\%);
\item derived new relative fractions of AGN and galaxies contributing to the observed MIR source counts, with AGN going from $\sim$10--20\% of the MIR population at $S_{15}<$0.6 mJy up to $\sim90-100$\% at $S_{15} > 10$ mJy;
\item computed new source counts in the MIR, with the AGN counts (especially type 2's) now increasing with a slope similar to that of galaxies. This result is likely to imply different evolutionary rates than formerly considered by the present models (i.e. lower rates for starburst galaxies and higher rates for type 2 AGN).
\end{itemize}

\acknowledgments
The authors acknowledge financial contribution from the contracts PRIN-INAF 1.06.09.05, ASI-INAF I/023/05/0 
and PRIN-MIUR 2006025203. MP acknowledges financial support from the Marie-Curie Fellowship grant 
MEIF-CT-2007-042111. We thank F. Fiore and M. Mignoli for useful discussions.  
We are grateful to the anonymous referee for useful comments that improved
the quality of the paper.

\clearpage


\pagestyle{empty}
\setlength{\voffset}{21mm}
\begin{deluxetable}{crrrrrrrrrrrrrrrccrr} 
\rotate
\tablewidth{0pt}
\tabletypesize{\tiny}
  \tablecaption{Multiwavelength Properties of the 15 $\mu$m Sources in the ELAIS-S1 Field: Example Table}
\label{tab_sed}
  \startdata
& & 
\\ \hline

ISOCAM NAME               &   F$_{FUV}$&   F$_{NUV}$&   F$_{B}$  &    F$_{V}$ &   F$_{R}$  &  F$_{J}$  &  F$_{Ks}$ &F$_{3.6}$&F$_{4.5}$&F$_{5.8}$&F$_{8.0}$&F$_{15}$&F$_{24}$& F$_{70}$&F$_{160}$& z & L$_{IR}$ & Spe$\_$cla & SED$\_$cla\\
                          & ($\mu$Jy) & ($\mu$Jy)   & ($\mu$Jy)  & ($\mu$Jy)  & ($\mu$Jy)  & ($\mu$Jy) & ($\mu$Jy) & ($\mu$Jy)    & ($\mu$Jy)    & ($\mu$Jy)    & ($\mu$Jy)    &   (mJy)    &     (mJy)   &    (mJy)     &    (mJy)  & & (L$_{\odot}$) & & \\ \hline
ELAISC15$\_$J002848$-$430658   &     8.4   &    26.7   &    43.3   &     $-$   &    39   &     $-$  &      $-$ &     301  &    413  &    560  &    780  &   1.0 &    2.4 &  $<$15.6 &  $<$99.6 & 0.594  & 8.88$\times 10^{12}$& AGN1&  AGN1\\
ELAISC15$\_$J002904$-$425243   &    15.7   &    40.1   &    56.6   &     $-$   &    67   &     $-$  &      $-$ &     537  &    841  &   1186  &   1706  &   1.2 &    3.8 &  $<$15.6 &  $<$99.6 & 0.642  & 1.15$\times 10^{13}$& AGN1&  AGN1\\
ELAISC15$\_$J002904$-$432415   &     $-$   &     $-$   &    69.4   &     $-$   &   152   &     $-$  &      $-$ &     223  &    200  &    121  &   1182  &   1.4 &    2.1 &    31.3 &   $<$99.6 & 0.207  & 1.99$\times 10^{12}$& GAL&  GAL\\
ELAISC15$\_$J002915$-$430333   &     $-$   &     $-$   &    21.3   &     $-$   &    70   &     $-$  &    356   &     297  &    352  &    281  &   1247  &   1.9 &    4.2 &   57.1  &  175.5  & 0.417   &  2.94$\times 10^{13}$& STB&  STB\\
ELAISC15$\_$J002924$-$432233   &    14.1   &    26.7   &    69.4   &     $-$   &   195   &     $-$  &    470   &     387  &    298  &    352  &   2517  &   2.1 &    3.8 &    49.3 &  $<$99.6  & 0.374  & 2.85$\times 10^{13}$& LINER&  AGN2\\
\enddata
\end{deluxetable}

\clearpage
\setlength{\voffset}{0mm}
\pagestyle{plaintop}

\begin{deluxetable}{lrlllll}
 \tablecaption{Spectroscopic Classification versus SED classification}
\tabletypesize{\small}
  \startdata
& & & & &  \\ \hline
 SED $\Downarrow$ $\backslash$ SPEC $\Rightarrow$ \hfill \vline& Tot~\vline & AGN1 & AGN2$+$LINER & Starburst& Galaxy & Unclass \\ \hline
 Tot \hfill \vline          & 203 \vline &   27~~~~~    &   25+7~~~~~~   &   32~~~~~~   &  107~~~~~~~~   &  5           \\ \hline
 AGN1 \hfill \vline         & 24  \vline &   20 (74.1\%)& ~1+0 (3.1\%)   &   ~2 (6.2\%) &  ~~~1 (0.9\%) &  0           \\
 AGN2  \hfill \vline        & 72  \vline &   ~3 (11.1\%)&  14+4 (56.3\%)~&   14 (43.8\%)& ~35 (32.8\%)  &  2 (40.0\%)   \\
 AGN$+$Starburst \hfill \vline& 11 \vline &   ~4 (14.8\%)& ~5+0 (15.6\%)  &   ~1 (3.1\%)&  ~~~1 (0.9\%)            &  0   \\
 Starburst \hfill \vline    & 13  \vline &    ~0~~~~~~  & ~2+1 (9.4\%)  &   ~6 (18.8\%)& ~~~3 (2.8\%)    &  1 (20.0\%)   \\
 Galaxy  \hfill \vline      & 83  \vline &    ~0~~~~~~  & ~3+2 (15.6\%)   &   ~9 (28.1\%)&  ~67 (62.6\%) &  2 (40.0\%)   \\  \hline
\enddata
\label{tab_class}
\end{deluxetable}

\clearpage


\begin{thebibliography}{}
\bibitem[Alonso-Herrero et al. 2006]{almudena06} Alonso-Herrero, A., et al. 2006, ApJ, 640, 167
\bibitem[{{Berta}(2005)}]{berta05}
{Berta}, S. 2005, Ph.D.~Thesis, University of Padua, Italy
\bibitem[Berta et al. 2006]{berta06} Berta, S., et al. 2006, A\&A, 451, 881
\bibitem[{{Berta} {et~al.}(2003){Berta}, {Fritz}, {Franceschini}, {Bressan}, \&
  {Pernechele}}]{berta03}
{Berta}, S., {Fritz}, J., {Franceschini}, A., {Bressan}, A., \& {Pernechele},
  C. 2003, \aap, 403, 119
\bibitem[Brand et al. 2006]{brand06} Brand, K., et al. 2006, \apj, 644, 143
\bibitem[{Brandt} \& {Hasingher} 2005]{brandt05} Brandt, W. N., \& Hasinger, G. 2005, ARA\&A, 43, 827
\bibitem[{{Bruzual} \& {Charlot}(2003)}]{bruzual03}
{Bruzual}, A.~G. \& {Charlot}, S. 2003, \mnras, 344, 1000
\bibitem[Burgarella et al. 2005]{burg05} Burgarella, D., et al. 2005, \apj, 619, L63
\bibitem[caccianiga et al. 2007]{caccianiga07} Caccianiga, A., Severgnini, P., Della Ceca, R., Maccacaro, T., Carrera, F.J., Page, M.J., 2007, A\&A, 470, 557
\bibitem[Cesarsky et al. 1996]{cesarsky96} Cesarsky, C.J., Abergel, A., Agn\`esel 
P., et al. 1996, A\&A, 315, L32
\bibitem[Chary and Elbaz 2001]{chary01} Chary, R., \& Elbaz, D. 2001, \apj, 556, 562
\bibitem[Chary et al. 2004]{chary04} Chary, R., et al. 2004, \apjs, 154, 80
\bibitem[Civano et al. 2007]{civano07} Civano, F., et al. 2007, A\&A, 476, 1223
\bibitem[Cocchia et al. 2007]{cocchia07} Cocchia, F., et al. 2007, A\&A 466, 31
\bibitem[{{Coleman} {et~al.}(1980){Coleman}, {Wu}, \& {Weedman}}]{coleman80}
{Coleman}, G.~D., {Wu}, C.-C., \& {Weedman}, D.~W. 1980, \apjs, 43, 393
\bibitem[Daddi et al. 2007]{daddi07} Daddi, E., et al. 2007, \apj, 670, 173
\bibitem[{{Devriendt} {et~al.}(1999){Devriendt}, {Guiderdoni}, \&
  {Sadat}}]{devriendt99}
{Devriendt}, J.~E.~G., {Guiderdoni}, B., \& {Sadat}, R. 1999, \aap, 350, 381
\bibitem[Elbaz et al. 1999]{elb99} Elbaz, D., et al. 1999, A\&A, 351, L37
\bibitem[Elbaz et al. 2002]{elb02} Elbaz, D., et al. 2002, A\&A, 384, 848
\bibitem[Farrah et al. 2003]{farrah03} Farrah, D., Afonso, J., Efstathiou, A., Rowan-Robinson, M., Fox, M. \& Clements, D. 2003, MNRAS, 343, 585
\bibitem[{{Fioc} \& {Rocca-Volmerange}(1997)}]{fioc97} {Fioc}, M., \& {Rocca-Volmerange}, B. 1997, \aap, 326, 950
\bibitem[Fiore et al. 2003]{fiore03} Fiore, F., et al. 2003, A\&A, 409, 79
\bibitem[Fiore et al. 2008]{fiore08} Fiore, F., et al. 2008, \apj, 672, 94
\bibitem[ForsterS et al. 2003]{fors03} F\"orster Schreiber, N.M., et al. 2003, A\&A, 399, 833
\bibitem[{Franceschini} {et~al.}(2005)]{franceschini05} {Franceschini}, A., {et~al.} 2005, \aj, 129, 2074
\bibitem[Franceschini et al. 2001]{fran01} Franceschini, A., et al. 2001, A\&A, 378, 1
\bibitem[Gruppioni et al. 2002]{cg02} Gruppioni, C., et al. 2002, MNRAS, 335, 831
\bibitem[Gruppioni et al. 2005]{cg05} Gruppioni, C., Pozzi, F., Lari, C., Oliver, S., and Rodighiero, G. 2005, ApJ, 618, L9
\bibitem[Hacking et al. 1987]{hack87} Hacking, P.B., Houck, J.R. \& Condon, J.J. 1987, ApJ, 316, 15
\bibitem[{{Hatziminaoglou} {et~al.}(2005){Hatziminaoglou}, {P{\'e}rez-Fournon},
  {Polletta}, {Afonso-Luis}, {Hern{\'a}n-Caballero}, {Montenegro-Montes},
  {Lonsdale}, {Xu}, {Franceschini}, {Rowan-Robinson}, {Babbedge}, {Smith},
  {Surace}, {Shupe}, {Fang}, {Farrah}, {Oliver}, {Gonz{\'a}lez-Solares}, \&
  {Serjeant}}]{hatziminaoglou05}
{Hatziminaoglou}, E., {P{\'e}rez-Fournon}, I., {Polletta}, M., {et~al.} 2005,
  \aj, 129, 1198
\bibitem[Hickox et al. 2007]{hickox07} Hickox, R.C., et al. 2007, ApJ, 671, 1365 
\bibitem[Hook et al. 1998]{hook98} Hook, I.M., Becker, R.H., McMahon, R.G., \& White, R.L. 1998, MNRAS, 297, 1115 
\bibitem[Ilbert et al. 2006]{ilbert06} Ilbert, O., et al. 2006, A\&A, 457, 841
\bibitem[Jarrett et al. 2000]{jarrett00} Jarrett, T.H., Chester, T., Cutri, R., Schneider, S., 
Skrutskie, M., \& Huchra, J.P. 2000, AJ 119, 2498
\bibitem[{Jarrett} {et~al.}(2006)]{jarrett06} {Jarrett}, T.~H., {et~al.} 2006, \aj, 131, 261
\bibitem[Kessler et al. 1996]{kessler96} Kessler M.F., et al. 1996, A\&A, 315, 27 
\bibitem[La Franca et al. 2004]{fabio04} La Franca, F., et al. 2004, AJ, 127, 3075
\bibitem[La Franca et al. 2007]{fabio07} La Franca, F., et al. 2007, A\&A, in press
\bibitem[Lagache et al. 1999]{lagache99} Lagache, G., Dole, H., \& Puget, J.-L. 1999, A\&A, 344, 322
\bibitem[Lagache et al. 2004]{lagache04} Lagache, G., et al. 2004, \apjs, 154, 112
\bibitem[Lari et al. 2001]{lari01} Lari, C., et al. 2001, MNRAS, 325, 1173
\bibitem[Le Floc$^{'}$h et al. 2005]{lefloch05} Le Floc$^{'}$h, E., et al. 2005, \apj, 632, 169 
\bibitem[Lonsdale and Hacking 1989]{lonsd89} Lonsdale, C.J., \& Hacking, P.B. 1989, ApJ, 339, 712
\bibitem[Lonsdale et~al. 2003]{lonsdale03} Lonsdale, C., {et~al.} 2003, PASP, 115, 897
\bibitem[Lonsdale et~al.2004]{lonsdale04} Lonsdale, C., {et~al.} 2004, \apjs, 154, 54
\bibitem[Lonsdale et~al. 2006]{lonsdale06} Lonsdale, C., Farrah, D. \& Smith, H.E., Astrophysics Update 2, 285
\bibitem[Maddox et al. 1990]{maddox90} Maddox, S.J., Efstathiou, G., Sutherland, W.J. \& Loveday, J. 1990, MNRAS, 243, 692
\bibitem[Maiolino et al. 2006]{maiolino06} Maiolino, R., et al. 2006, A\&A 445, 457
\bibitem[Marleau et al. 2004]{marleau04} Marleau, F., et al. 2004, \apjs, 154, 66
\bibitem[Martin et al. 2005]{martin05} Martin D.C., et al. 2005, \apj, 619, L1
\bibitem[Martinez-Sansigre et al. 2007]{alejo07} Martinez-Sansigre, A., \& Rawlings, S. 2007, proc. of 
``The Central Engine of Active Galactic Nuclei''. Eds. L.C. Ho \& J.-M. Wang (San Francisco: ASP), v. 373, p. 728
\bibitem[Matute et al. 2002]{matute02} Matute, I., et al. 2002, MNRAS, 332, L11
\bibitem[Matute et al. 2006]{matute06} Matute, I., La Franca, F., Pozzi, F., Gruppioni, C., Lari, C., \& Zamorani, G. 2006, A\&A, 451, 553
\bibitem[Papovich et al. 2004]{papovich04} Papovich, C., et al. 2004, \apjs, 154, 70
\bibitem[Pearson 2005]{pearson05} Pearson C. 2005, MNRAS, 358, 1417
\bibitem[Perez-Gonzalez 2005]{perezgonzalez05} P\`erez-Gonz\`alez, P.G., et al. 2005, \apj, 630, 82
\bibitem[{Polletta} {et~al.}(2006)]{polletta06} Polletta, M., et~al. 2006, \apj, 642, 673
\bibitem[{Polletta} {et~al.}(2007)]{polletta07} {Polletta}, M., et al. 2007, \apj, 663, 81 
\bibitem[Pozzi et al. 2004]{pozzi04} Pozzi, F., et al. 2004, ApJ, 609, 122
\bibitem[Puccetti et al. 2006]{puccetti06} Puccetti, S., et al. 2006, A\&A, 457, 501
\bibitem[Rowan-Robinson et al. 2004]{mrr04} Rowan-Robinson, M., et al. 2004, MNRAS, 351, 1290 
\bibitem[Rush et al. 1993]{rush93} Rush, B., Malkan, M.A., \& Spinoglio, L. 1993, \apjs ,89, 1
\bibitem[Saunders et al. 1990]{saund90} Saunders, W., Rowan-Robinson, M., Lawrence, A., Efstathiou, G., Kaiser, N.,
Ellis, R.S., \& Frenk, C.S. 1990, MNRAS, 242, 318
\bibitem[{Silva} {et~al.}(1998)]{silva98} {Silva}, L., {Granato}, G.~L., {Bressan}, A., \& {Danese}, L. 1998, \apj, 509, 103
\bibitem[Smith et al. 2007]{smith07} Smith, J.D.T., et al. 2006, ApJ, 656, 770
\bibitem[Stern et al. 2005]{stern05} Stern, D., et al. 2005, ApJ, 631, 163
\bibitem[Surace et al. 2005]{surace05} Surace, J.A., et al. 2005, AAS, 207, 37, p.1246 
\bibitem[{Tajer} {et~al.}(2007)]{tajer07}
{Tajer}, M., {et~al.} 2007, A\&A, 467, 73 
\bibitem[Treister et al. 2006]{treister06} Treister, E., et al., 2006, ApJ, 640, 603
\bibitem[Vanzella et al. 2005]{vanzella05} Vanzella, E., et al. 2005, A\&A, 434, 53 
\bibitem[Vanzella et al. 2006]{vanzella06} Vanzella, E., et al. 2006, A\&A, 454, 423
\bibitem[{Weedman} {et~al.}(2006)]{weedman06a}
{Weedman}, D., et al. 2006, \apj, 653, 101
\bibitem[Werner et al. 2004]{werner04} Werner, M.W., et al. 2004, \apjs, 154, 1

\end{thebibliography}
\end{document}